\documentclass{article}
\usepackage{pgf-umlsd}
\usepackage{listings}
\usepackage{csquotes}
\usepackage{fancyhdr}

\usepackage[numbers]{natbib}
\usepackage{tikz}
\usetikzlibrary{shapes, arrows, positioning, fit, calc}

\makeatletter
\if@titlepage
  \renewenvironment{abstract}{%
      \titlepage
      \null\vfil
      \@beginparpenalty\@lowpenalty
      \par\medskip\noindent{\bfseries\abstractname:}
        \@endparpenalty\@M
      }%
     {\par\vfil\null\endtitlepage}
\else
  \renewenvironment{abstract}{%
      \if@twocolumn
        {\bfseries\abstractname:}%
      \else
        \quotation
        \small\noindent{\bfseries\abstractname:}
        \@endparpenalty\@M
      \fi}
      {\if@twocolumn\par\medskip\else\endquotation\fi}
\fi
\makeatother

\title{Universal Session Protocol \\ 
\large Mitigating Unauthenticated Remote Code Execution
}
\author{Jonathon Anderson \\ \small \texttt{jonathon.anderson@protonmail.com}}
\date{}

\begin{document}
  \maketitle
  \begin{abstract}
    Currently, the TCP/IP model enables exploitation of vulnerabilities anonymously by unconditionally fulfilling every request for a connection into an application; the model only incorporates authentication within applications themselves, rather than as a precondition for access into applications. I am proposing the Universal Session Protocol as a change to the architecture of the TCP/IP model to include a session layer featuring a structured generalized process for authentication negotiation and fulfillment. The Universal Session Protocol addresses an urgent and vital need to eliminate unauthenticated data processing on security critical systems. Previous work regarding TCP/IP security has focused on the application design and implementation and existing protocol layers, but has failed to posit the addition of a session layer as a mitigating control. Failing to implement a distinct authentication layer leaves every resource connected to the global Internet, including life and security critical infrastructure, vulnerable to attacks from anonymous and untraceable sources. The Universal Session Protocol provides a solution by establishing a TCP/IP Session Layer that explicitly provides authentication before a data stream is accessible within an application. After authentication, an identity is associated with the data stream so that all data may be related back to that identity for forensic purposes. If authentication fails, the application will never process user data, rendering the service safe from anonymous bad actors. 
  \end{abstract}

  \section{Introduction}
  \begin{displayquote}
      \begin{center}
        Vanquish your foes by always keeping yourself in a safe and unassailable position; then no one will suffer any losses.

        Morihei Ueshiba 
      \end{center}
    \end{displayquote}
  Anonymity is simultaneously a great benefit and challenge inherent to the public Internet environment. With minimal effort and technical know-how, whistle blowers, political dissidents, and the general public can be assured that their identity is virtually untraceable by would be persecutors. Attempting to architecturally close the security vulnerability enabling anonymous exploitation by requiring universal authentication for internet facing applications is untenable even before the thought occurs. Consequently, bad actors enjoy the same benefits, much to the frustration of system developers, administrators, and security teams. Because there is no generalized authentication mechanism in the TCP/IP model, applications on the internet are responsible for establishing their own authentication schemes, building their systems on top of standardized application layer protocols. The applications that process these protocols process every byte initially the same, regardless of source. The best applications, identity, and access systems are plagued by the reality that, in some form or another, users are interacting with the application responsible for processing the base protocol prior to establishing an identity. As such, all data must be allowed in for initial processing, and vulnerabilities in the core application, or any internal or extensible feature, can be exploited with little recourse; tracking skilled anonymous actors can require a major concerted effort and is often impossible. In any case, by the time a search has begun, the damage has been done.

  The Universal Session Protocol (USP) seeks to fulfill both objectives of permitting anonymity for sensitive or everyday scenarios and eliminating abuse of applications. The technique itself is simple, and the protocol aims to be as lightweight as possible. Rather than a TCP/IP stream being created directly by the application, the session layer abstraction uses the stream to exchange connection establishment messages, and, authenticating if necessary, only then providing a stream to the application. A connection not requiring authentication transmits one message in each direction to establish a session, while a connection requiring authentication transmits five messages total, with additional overhead for the authentication protocol itself. Once authentication is complete, the stream is passed to the application along with an identity token so that each and every byte sent into the application may be associated with that identity. If authentication fails, the stream is disconnected, and no data is processed within the application, rendering any attacks on the application impossible.

  \section{Related Work}
    Since the inception of TCP/IP, there have been no known efforts to explore the implications of the design choice regarding the use or exclusion of a generic session layer protocol. Because all OSI Session Layer functionality has been implemented within the TCP/IP Application Layer, there may be little benefit to considering whether that functionality should be implemented in a generic session layer. As we can see, the internet is working well enough that there really is no need to question the model. However, authentication is different because guaranteed identification of what data is processed is only possible by implementing a protocol that establishes identities before the application processes the first byte. Currently, authentication data and processes are handled in the application layer and therefore treated as application data and processes. Because the application layer does not internally dictate an order of processing, application data may be and is processed before authentication and identity data. Authentication data and processes must be treated as separate and distinct entities prior to application data processing to ensure that all application data is properly associated with an identity.

    As such, the sole purpose of this paper is to establish the need for and definition of the Universal Session Protocol on the grounds of enhanced security. While it may turn out to be sensible to implement additional functionality in the session layer, that will be the subject of future research and proposals. For the purpose of this current proposal, the security merits stand on their own and are better served by highlighting them exclusively. A brief review of the literature demonstrates that security research for TCP/IP is limited to improving program design and implementation in the application layer as well as design and implementation of the physical, data link, network, and transport layers.

    Perhaps the most authoritative source of knowledge regarding application security is the Online Web Application Security Project (OWASP), founded in 2001 with the objective to have no insecure applications\cite{owasp}. While the efforts of OWASP have been generally successful and informative, the methods and designs do not address the issue of treating authentication data as application data. If it turns out that, instead of application or identity data, a malicious payload was transmitted, the payload will be processed as application data without first establishing identity. It is only by treating identity and authentication data as a distinct category of data processed during a distinct phase of session establishment that we can ensure that application data is only processed after an identity has been established. This method restricts access to all vulnerabilities cataloged by OWASP to only connections with established identities.

	  Besides application layer vulnerabilities, much research has been conducted into the TCP/IP protocol layers below the application layer: physical, data link, network, and transport. The nature of this research focuses on vulnerabilities that result from the architecture and design of TCP/IP such as information theft, denial of service, spoofing, sniffing, session hijacking, fragmentation attacks, and information destruction\cite{harris}\cite{bellovin}. While some papers did identify IP based authentication as a major flaw, the proposed solutions stopped short of suggesting a dedicated authentication layer\cite{bellovinrevisited}. The only research that was found to have suggested the addition of a session layer was strictly speaking in terms of functional session management requirements and failed to mention authentication as a concern\cite{landfeldt}.

  \section{Connection Security Flow}
  \subsection{Current Model}
  
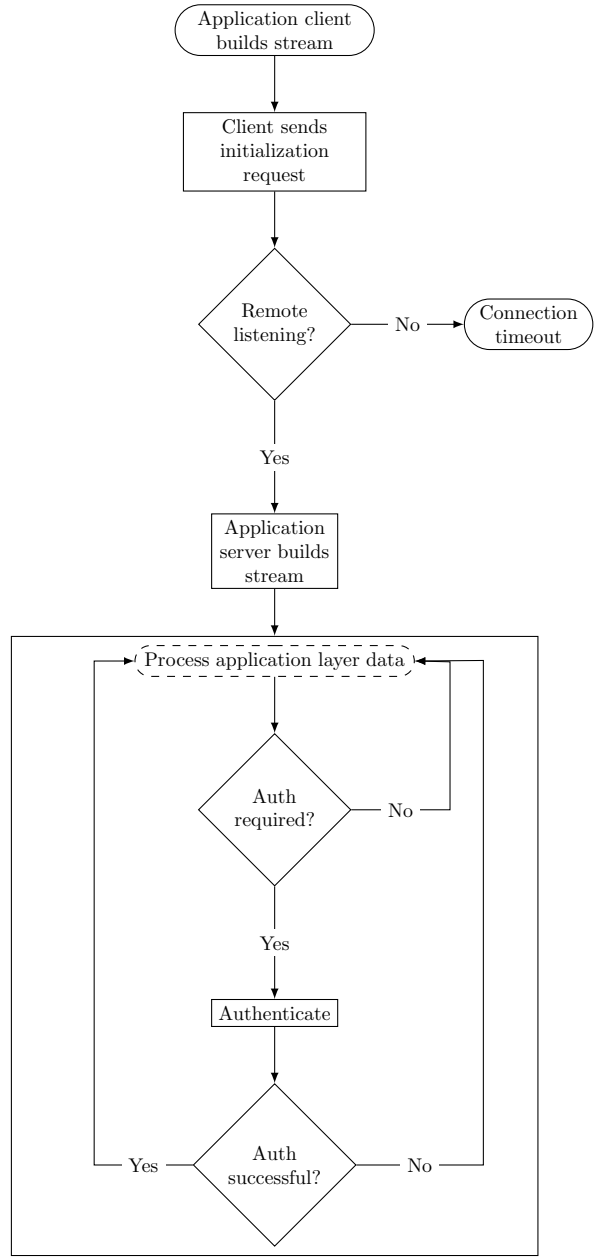
\begin{center}
    \begin{tikzpicture}
      \begin{scope}[scale=0.75, transform shape]
        \node[draw, 
          rounded rectangle,
          align=center,
          text width=3.0cm]
          (start)
          {Application client builds stream};
        \node[draw, 
          rectangle,
          text width=3cm,
          align=center,
          below=of start]
          (client_stream)
          {Client sends\\ initialization request};
        \node[draw, 
          diamond,
          text width=1.5cm,
          align=center,
          below=of client_stream]
          (remote_listening)
          {Remote listening?};
        \node[draw,
          rounded rectangle,
          text width=1.75cm,
          align=center,
          right=2cm of remote_listening]
          (timeout)
          {Connection\\timeout};
        \node[draw,
          rectangle,
          text width=2cm,
          align=center,
          below=2cm of remote_listening]
          (server_stream)
          {Application server builds stream};
        \node[draw,
          dashed,
          rounded rectangle,
          align=center,
          below=of server_stream]
          (process_data)
          {Process application layer data};
        \node[draw,
          diamond,
          text width=1.5cm,
          align=center,
          below=of process_data]
          (auth_required)
          {Auth required?};
        \node[draw,
          rectangle,
          text width=2cm,
          align=center,
          below=2cm of auth_required]
          (authenticate)
          {Authenticate};
        \node[draw,
          diamond,
          text width=1.75cm,
          align=center,
          below=of authenticate]
          (auth_successful)
          {Auth successful?};
        \node[fit=(process_data) (auth_required) (authenticate) (auth_successful),
          draw,
          rectangle,
          align=center,
          minimum width=7cm]
          (application)
          {};

          \draw[-latex] (start)         edge (client_stream)
                        (client_stream) edge (remote_listening)
                        (server_stream) edge (application)
                        (authenticate)  edge (auth_successful)
                        (process_data)  edge (auth_required);
          \draw[-latex] (remote_listening) edge node[fill=white]{No}  (timeout);
          \draw[-latex] (remote_listening) edge node[fill=white]{Yes} (server_stream);
          \draw[-latex] (auth_required.east) -- node[fill=white]{No} +(1.75,0) -- +(1.75,2.625) -- (process_data.east);
          \draw[-latex] (auth_required) edge node[fill=white]{Yes} (authenticate);
          \draw[-latex] (auth_successful.west) -- node[fill=white]{Yes} +(-1.75,0) -- +(-1.75,8.94) -- (process_data.west);
          \draw[-latex] (auth_successful.east) -- node[fill=white]{No} +(2.25,0) -- +(2.25,8.95) -- (process_data.east);
      \end{scope}
    \end{tikzpicture}
  \end{center}

    Application layer data is always processed before authentication occurs. The authentication process and authentication data are handled as application data.
    \newpage

  \subsection{Proposed Model}
  
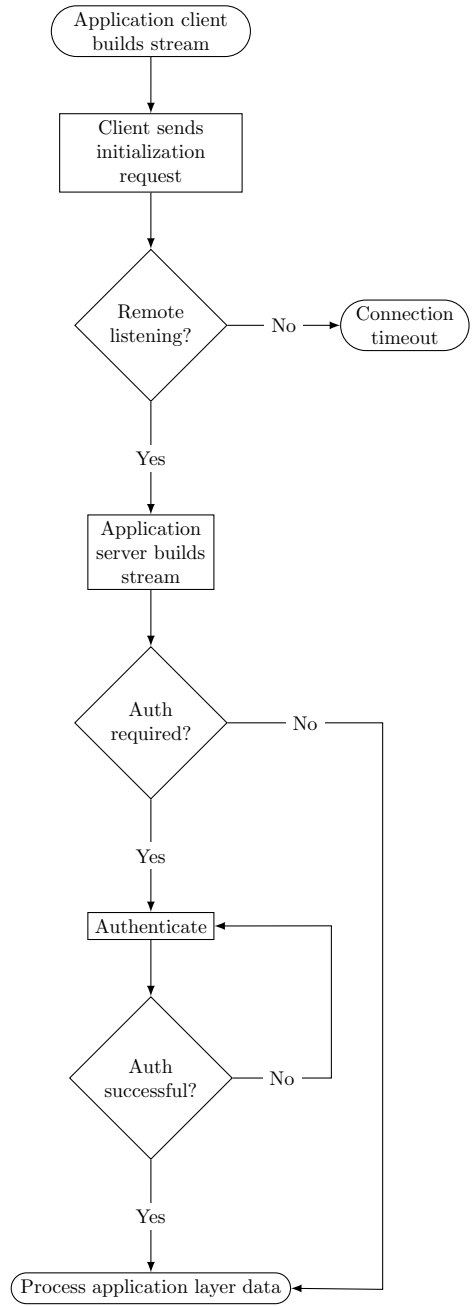
\begin{center}
    \begin{tikzpicture}
    \begin{scope}[scale=0.75, transform shape]
      \node[draw, 
        rounded rectangle,
        align=center,
        text width=3.0cm]
        (start)
        {Application client builds stream};
      \node[draw, 
        rectangle,
        text width=3cm,
        align=center,
        below=of start]
        (client_stream)
        {Client sends\\ initialization request};
      \node[draw, 
        diamond,
        text width=1.5cm,
        align=center,
        below=of client_stream]
        (remote_listening)
        {Remote listening?};
      \node[draw,
        rounded rectangle,
        text width=1.75cm,
        align=center,
        right=2cm of remote_listening]
        (timeout)
        {Connection\\timeout};
      \node[draw,
        rectangle,
        text width=2cm,
        align=center,
        below=2cm of remote_listening]
        (server_stream)
        {Application server builds stream};
      \node[draw,
        diamond,
        text width=1.5cm,
        align=center,
        below=of server_stream]
        (auth_required)
        {Auth required?};
      \node[draw,
        rectangle,
        text width=2cm,
        align=center,
        below=2cm of auth_required]
        (authenticate)
        {Authenticate};
      \node[draw,
        diamond,
        text width=1.75cm,
        align=center,
        below=of authenticate]
        (auth_successful)
        {Auth successful?};
      \node[draw,
        rounded rectangle,
        align=center,
        below=2cm of auth_successful]
        (process_data)
        {Process application layer data};

        \draw[-latex] (start)         edge (client_stream)
                      (client_stream) edge (remote_listening)
                      (server_stream) edge (auth_required)
                      (authenticate)  edge (auth_successful);
        \draw[-latex] (remote_listening) edge node[fill=white]{No}  (timeout);
        \draw[-latex] (remote_listening) edge node[fill=white]{Yes} (server_stream);
        \draw[-latex] (auth_required.east) -- node[fill=white]{No} +(2.75,0) -- +(2.75,-10.05) -- (process_data.east);
        \draw[-latex] (auth_required) edge node[fill=white]{Yes} (authenticate);
        \draw[-latex] (auth_successful) edge node[fill=white]{Yes} (process_data);
        \draw[-latex] (auth_successful.east) -- node[fill=white]{No} +(1.75,0) -- +(1.75,2.7) -- (authenticate);
    \end{scope}
    \end{tikzpicture}
  \end{center}

  Authentication will always occur before application layer data is processed. The authentication process and authentication data are handled separately.
  \newpage

  \section{Control Flow}
  USP can be implemented by either the operating system or the application being served, but the control flows are the same. Message descriptions are provided after the control flow sequence diagrams. Implementation details are addressed in the Discussion section.
  
  \vbox{
    \subsection{Unauthenticated Connection - Direct}
    Establishing a connection between client and server that does not require authentication is fast and simple. The client sends a request to initialize an application session, the server determines that the requested application does not require authentication, and then sends a message for the client to complete the connection.

    Because the stream is already built and active, connecting is effortless; both server and client simply return or pass the existing stream into the target processes.

      
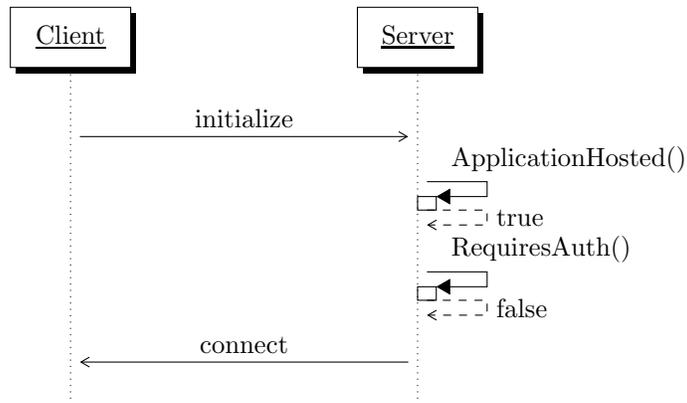
\begin{center}
        \begin{sequencediagram}
          \newinst{client}{Client}
          \newinst[3]{server}{Server}

          \mess{client}{initialize}{server}
          \begin{call}{server}{ApplicationHosted()}{server}{true}\end{call}
          \begin{call}{server}{RequiresAuth()}{server}{false}\end{call}
          \mess{server}{connect}{client}
        \end{sequencediagram}
      \end{center}
  }
    
      \vbox{
        \subsection{Authenticated Connection - Direct}
        It's unavoidable that coordinating and completing authentication will incur additional overhead. That being said, the authentication overhead is minimal and message sizes are small. The final connection process after authentication uses the same two-message model for unauthenticated connections: an initialize message is sent with valid tokens, and a connect message from the server completes the process. Any authentication protocol may be used. 

          \begin{center}
            \begin{sequencediagram}
              \newinst{client}{Client}
              \newinst[0.5]{client_auth}{Auth}
              \newinst[1]{server_auth}{Auth}
              \newinst[0.5]{server}{Server}

              \mess{client}{initialize}{server}
              \begin{call}{server}{ApplicationHosted()}{server}{true}\end{call}
              \begin{call}{server}{RequiresAuth()}{server}{true}\end{call}
              \begin{call}{server}{NegotiateAuthProtocol()}{server}{true}\end{call}
              \begin{call}{server}{TokenValid()}{server}{false}\end{call}
              \mess{server}{authenticate}{client}

              \begin{call}{server}{beginAuth()}{server_auth}{success}
                \begin{call}{client}{beginAuth()}{client_auth}{success}
                  \begin{call}{client_auth}{credentials}{server_auth}{success}
                  \end{call}
                \end{call}
              \end{call}

              \mess{server}{token}{client}
              \mess{client}{initialize}{server}
              \begin{call}{server}{ApplicationHosted()}{server}{true}\end{call}
              \begin{call}{server}{RequiresAuth()}{server}{true}\end{call}
              \begin{call}{server}{NegotiateAuthProtocol()}{server}{true}\end{call}
              \begin{call}{server}{TokenValid()}{server}{true}\end{call}
              \mess{server}{connect}{client}

            \end{sequencediagram}
          \end{center}
      }

      \vbox{
        \subsection{Authenticated Connection - Token Passing}
        Because the final connection process for an authenticated connection uses the initialize/connect pattern, connections can be completed in two steps if a token was acquired by some other method or at a different time. This scenario is especially interesting in environments where a user should only be granted access under the discretion of other users or access control systems and should not be allowed to initiate sessions independently.

        
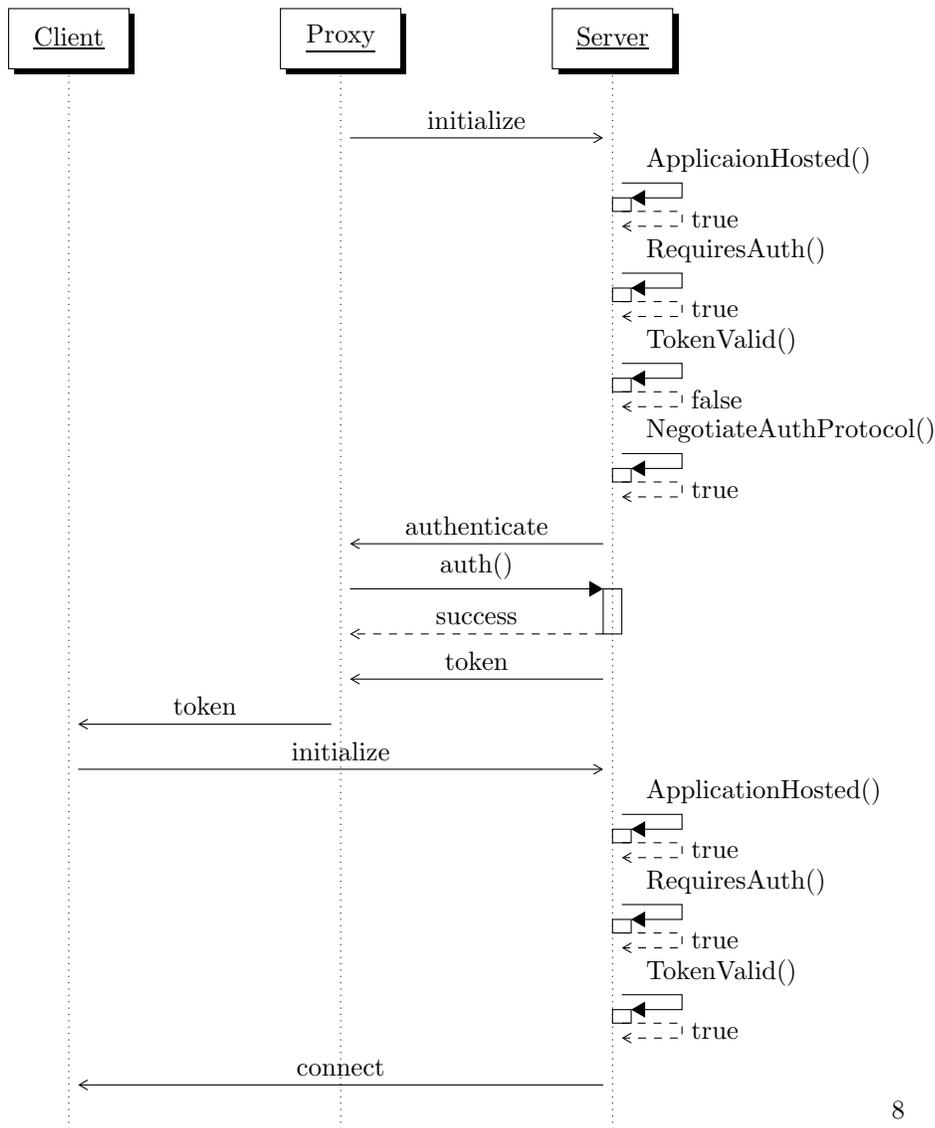
\begin{center}
          \begin{sequencediagram}
            \newinst{client}{Client}
            \newinst[2]{proxy}{Proxy}
            \newinst[2]{server}{Server}

            \mess{proxy}{initialize}{server}
            \begin{call}{server}{ApplicaionHosted()}{server}{true}\end{call}
            \begin{call}{server}{RequiresAuth()}{server}{true}\end{call}
            \begin{call}{server}{NegotiateAuthProtocol()}{server}{true}\end{call}
            \begin{call}{server}{TokenValid()}{server}{false}\end{call}
            
            \mess{server}{authenticate}{proxy}
            \begin{call}{proxy}{auth()}{server}{success}\end{call}

            \mess{server}{token}{proxy}
            \mess{proxy}{token}{client}
            \mess{client}{initialize}{server}
            \begin{call}{server}{ApplicationHosted()}{server}{true}\end{call}
            \begin{call}{server}{RequiresAuth()}{server}{true}\end{call}
            \begin{call}{server}{NegotiateAuthProtocol()}{server}{true}\end{call}
            \begin{call}{server}{TokenValid()}{server}{true}\end{call}
            \mess{server}{connect}{client}

          \end{sequencediagram}
        \end{center}
      }

      \vbox{
        \subsection{Authenticated Connection - No Shared Protocol}
        Because the initialization message provides information about what authentication protocols are supported by the client, the server may negotiate authentication protocols locally without additional data transmissions. If the client and server do not share an authentication protocol, an error message is sent from the server to the client.

        
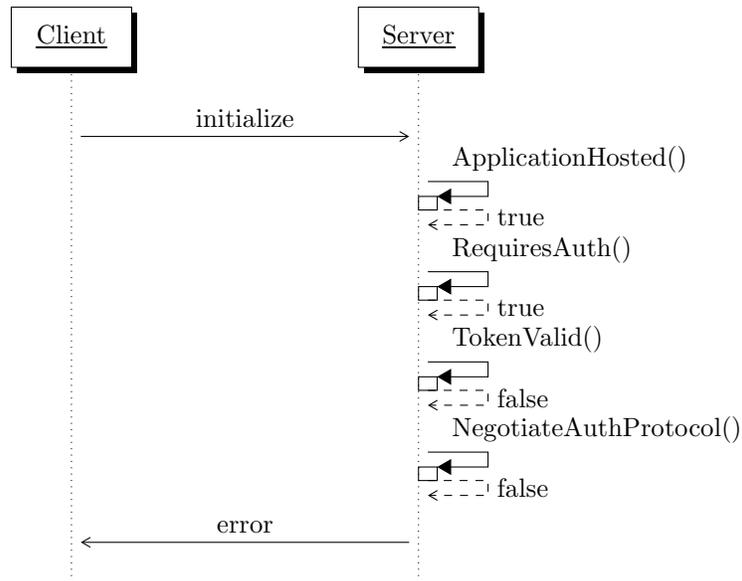
\begin{center}
          \begin{sequencediagram}
            \newinst{client}{Client}
            \newinst[3]{server}{Server}

            \mess{client}{initialize}{server}
            \begin{call}{server}{ApplicationHosted()}{server}{true}\end{call}
            \begin{call}{server}{RequiresAuth()}{server}{true}\end{call}
            \begin{call}{server}{NegotiateAuthProtocol()}{server}{false}\end{call}
            \begin{call}{server}{TokenValid()}{server}{false}\end{call}
            \mess{server}{error}{client}

          \end{sequencediagram}
        \end{center}
      }

      \vbox{
        \subsection{Authenticated Connection - Authentication Failure}
        Because the authentication process itself will return a success or failure, there is no need for an error message sent from server to client; the client and server both infer the status of authentication and terminate the session when a failure is reported locally to each system.

          
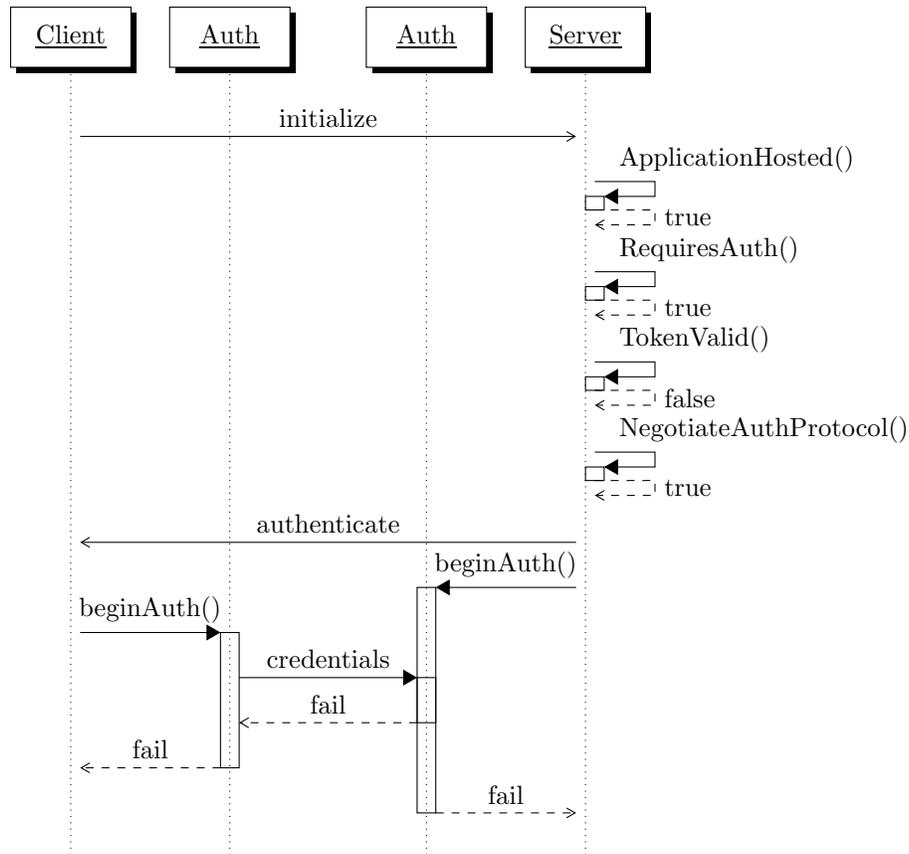
\begin{center}
            \begin{sequencediagram}
              \newinst{client}{Client}
              \newinst[0.5]{client_auth}{Auth}
              \newinst[1]{server_auth}{Auth}
              \newinst[0.5]{server}{Server}

              \mess{client}{initialize}{server}
              \begin{call}{server}{ApplicationHosted()}{server}{true}\end{call}
              \begin{call}{server}{RequiresAuth()}{server}{true}\end{call}
              \begin{call}{server}{NegotiateAuthProtocol()}{server}{true}\end{call}
              \begin{call}{server}{TokenValid()}{server}{false}\end{call}
              \mess{server}{authenticate}{client}

              \begin{call}{server}{beginAuth()}{server_auth}{fail}
                \begin{call}{client}{beginAuth()}{client_auth}{fail}
                  \begin{call}{client_auth}{credentials}{server_auth}{fail}
                  \end{call}
                \end{call}
              \end{call}
              
              \mess{server}{error}{client}

            \end{sequencediagram}
          \end{center}
      }

      \vbox{
        \subsection{Application Not Hosted}
        In the case that a client sends an initialize message requesting an application that is not hosted by the server, the server will return an error message.

          
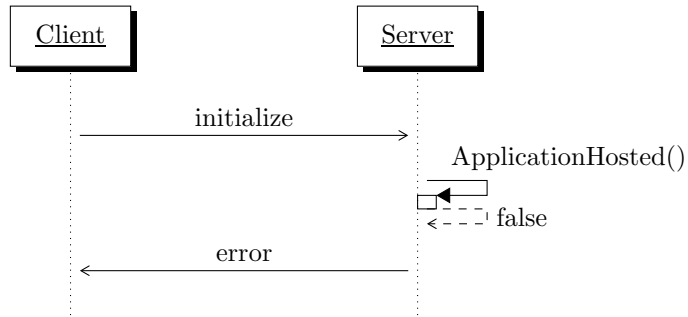
\begin{center}
            \begin{sequencediagram}
              \newinst{client}{Client}
              \newinst[3]{server}{Server}

              \mess{client}{initialize}{server}
              \begin{call}{server}{ApplicationHosted()}{server}{false}\end{call}
              \mess{server}{error}{client}]

            \end{sequencediagram}
          \end{center}
      }

      \vbox{
        \subsection{Malformed Message}
        In every message exchange, the receiving system will first ensure that the data structure is correct, thereby mitigating attacks that may take advantage of processing on malformed data.

          \begin{center}
            \begin{sequencediagram}
              \newinst{sender}{Sender}
              \newinst[3]{recipient}{Recipient}

              \mess{sender}{message}{recipient}
              \begin{call}{recipient}{ValidMessageStructure()}{recipient}{false}\end{call}
              \mess{recipient}{error}{client}]

            \end{sequencediagram}
          \end{center}
      }

  \newpage
  \section{Messages}
  \vbox{
    \subsection{initialize}
    \begin{lstlisting}
      {
        "message":"initialize",
        "authentication":["<protocol>", "<protocol>"],
        "streams": [
          {
            "application":"<application>",
            "token":"<token>"
          }
        ]
      }
    \end{lstlisting}
  }

  \vbox{
    \subsection{connect}
    \begin{lstlisting}
      {
        "message":"connect",
        "application":"<application>"
      }
    \end{lstlisting}
  }

  \vbox{
    \subsection{authenticate}
    \begin{lstlisting}
      {
        "message":"authenticate",
        "protocol":"<protocol>"
      }
    \end{lstlisting}
  }

  \vbox{
    \subsection{token}
    \begin{lstlisting}
      {
        "message":"token",
        "streams": [
          {
            "application":"<application>",
            "token":"<token>"
          }
        ]
      }
    \end{lstlisting}
  }

  \vbox{
    \subsection{error}
    \begin{lstlisting}
      {
        "message":"error",
        "error":"<text>"
      }
    \end{lstlisting}
  }

  \section{Discussion}
    \subsection{Implementation: Operating System vs Application}
    USP can be implemented by applications themselves, taking advantage of the existing network stack as is, or it can be implemented by operating systems, with the potential for subtle but dramatic changes to our internet infrastructure.

      \subsubsection{Application}
      Because applications are capable of and responsible for binding sockets by making system calls directly, they have the freedom to implement USP as an abstraction layer between stream allocation and the application's access to that stream. This is distinct from developing authentication into the hosted application because the stream resource is unavailable to the application until authentication has been completed. By contrast, traditional authentication relies on an application layer protocol to transmit authentication data back and forth, meaning that the stream and anonymous data are already being processed inside the application in full, even before determining if there is an identity to consider. Inserting a USP library drastically increases security by preventing the processing of unauthenticated data. Because of some design opportunities available by implementing USP in the operating system, developing the protocol into applications is probably the first route to initial adoption, as it does not require changes to current protocols and models that may conflict with the ability to route packets through networks.

      \subsubsection{Operating System}
      One interesting property of the TCP/IP model, compared to the OSI model, is the elimination of the independent session and presentation layer. A consequence of this is the requirement to statically map some transport layer identifier, i.e. ports, to some process in order for the system to know where to direct newly crafted streams and stream data from incoming connections, and these new connections are always accepted. There is, essentially, still a session layer in the TCP/IP model, but it is static and critically offers no authentication. USP, by contrast offers a programmatic mechanism to establish streams for applications without a previously known port by matching on the name of the application layer protocol and features authentication mechanisms to prevent unwanted data streams. The current model involves an ongoing competition for arbitrarily chosen port numbers that must be manually configured and updated according to whatever works best in that circumstance. This form of session management requires developers of applications to embed or enable configuration of transport layer information into the application rather than simply specifying what application is being offered and relying on a session layer abstraction to handle the mapping of data streams to applications. The current model universally requires prior knowledge of the configuration state of the target system and necessitates outside channels, such as documentation on websites or standardization, to communicate that information. This turns out to be unnecessarily inefficient compared to what computers are capable of and do on a constant basis: exchange information. Changing the way transport layer addresses are exchanged during session establishment may alleviate this inefficiency and remove port configurations as a requirement for application management. Finally, the current model leads to a finite address space because the port range is limited to the also static and previously agreed on standards.

      If we take the abstractions of the network stack to their ideal conclusion, the objective becomes accessing a remote service with only two pieces of information: an application name and a host name. With USP, it is possible for the transport layer and application layer each to be agnostic of the other. If the operating system provides a USP agent, then all streams established by the transport layer will be directed to their respective process, by USP, based on the application specified in the request and the application declared by the service. Routing the stream to the process is dynamic, no prior knowledge of a target system is required, there is no need to establish typical patterns for port mappings, and session negotiation happens explicitly. Most importantly, an authentication mechanism is introduced that ensures identity is established before processing any user data. Routing streams into applications can be as simple as returning the stream from a function call made within the application, or as sophisticated as using inter-process communication to deliver a stream without request from the service, perhaps even launching the requested application in the process. The host IP becomes a singular target and, instead of addressing remote hosts with port numbers, port numbers are exchanged as additional information during connection establishment. That is, the client will establish a local port number and send it to the server in the connection request, without a destination port number for the server. The server will then establish its own local port number and send that to the client in the connection response, addressing the port number received in the connection request. Both hosts will then identify the stream, as usual, by local and remote IP and port pairs and, once established, the USP agent determines which application, if any, will receive it.

    \subsection {Authentication-type specific token validation}
    Negotiating the authentication protocol before validating the token allows for an authentication-type specific validation. Both the authentication function and token validation function may be passed in as parameters to facilitate caller supplied functions, meaning that authentication information is no longer part of the application data stream, and the interface can be built directly into the user application.

    \subsection{Encryption and PKI} 
    Because USP has first access to the data stream, it can implement encryption as well as server identity verification, giving a ready packaged secure communication channel back to the application.

    \subsection{Anonymous Exploitation of Vulnerabilities}
    Because authentication and authorization must occur before a stream is passed into a process, scanning applications for and exploiting vulnerabilities becomes either impossible, if authentication fails, or traceable, if authentication succeeds. Unless an attacker is impersonating a legitimate user, they are conducting reconnaissance and attempting exploits at the peril of exposing themselves; there would be no anonymous exploitation of higher layer protocols or applications. At the very least, identifying the actor enables suspension of the account, protecting the application from further attack. Depending on what data is collected, how thoroughly it is verified during account registration, and what avenues are available for recourse, the risk of additional consequences may further serve as a deterrent. Even in the case of impersonation, having an identity to investigate is a starting point for determining how the identity was compromised and educating users or implementing additional safeguards.

    \subsection{Firewalls}
    Misconfiguration of firewalls is a significant security and availability concern. In firewall management, the requirement is to discern what has been configured on the target system and correlate this with lists of known and unknown addresses attempting to access the application, which may not always represent the same entity or person. This entangles network, transport, session, and application layer concerns in a way that complicates configuration, leads to security and access errors, and makes impossible full separation of administrative responsibilities. Implementing USP changes the patterns for access management from address based to identity based, relieving network engineers of the enormous and complex burden of trying to manage access using fungible addresses representing unknown actors. To be sure, firewalls could still implement IP blocking as an adjunct to access management, but there would be no requirement to, and no port driven configurations to manage, which significantly reduces complexity and opportunity for errors.

    \subsection{Isolation of Public and Secure Services}
    In the current model, web applications may host a public landing page and the secure elements of an application within the same process. This will no longer be possible with USP because the connection for application data itself is either authenticated or not, rather than being determined on a packet by packet basis. This will lead to the requirement that public facing elements of an application are hosted in a separate process with different USP authentication requirements. Ultimately, this improves the security of the application by preventing unauthenticated data from being processed alongside authenticated data.

    \subsection{Authentication Protocol Support}
    Because the application used by the end user is interacting directly with the USP abstraction layer, USP can be designed to accept pointers to functions that complete the authentication process for various protocols, using the application itself to interact with the user.

    \subsection{Passing Identity}
    Passing identity information between the session and application layer is possible because of the adjacency of the layers in the protocol stack. If the USP agent passes the tokens created during authentication to the hosting application, the application can use that token for identity driven access controls and logging. A client can request a token for itself for future use, or it can be requested by a third party proxy and transferred freely. This enables scenarios where a user may need legitimate access to a system, but only under the discretion of authorizing agents and other control systems, or may want to reserve system access for a future time. In both scenarios, the token is simply passed into the client USP agent and included in an "initialize" message. The client could also require the server to authenticate and record logs itself. 

    Current authentication schemes related to establishing identity within the TCP/IP stack, such as IPSec, cannot share this information with application layer protocols because there is no mechanism for passing identity information through the transport layer so that the network and application layer can share that data.

    \subsection{URL Formatting and Application Naming}
    Existing URL formats support USP because they contain the two pieces of information needed to access a service: the application name and the remote host name.

  \section{Further Research}
    \subsection{On-Demand Execution}
    USP could provide a novel approach to remote management of services by implementing the USP agent such that it can start, stop, and manage hosted applications. The message content could include a "payload" property that is simply properly formatted JSON. The JSON payload could then be attached to the stream object that is passed to the application, providing parameters for the execution of the application.

    \subsection{USP Agent Dual Role}
    Additional features are possible if the USP agent is implemented so that hosts function as both USP server and client. 

      \subsubsection{Full Duplex Communication}
      Full duplex command and control during active sessions would maximize flexibility. This may be possible by creating two command and control streams.
    
      \subsubsection{Third Party Brokering}
      If a target machine hosts an application that is intended to connect to a third machine and complete some kind of application work, a control server could send a USP message instructing the target machine to both initialize the application itself, using on-demand execution, and create a stream to the third machine, possibly initializing the application on that host as well. 
      
      \subsubsection{Management of Active Sessions}
      In its current description, USP is strictly an authentication protocol. There is potential for more sophisticated session management if command messages can be exchanged bidirectionally while the streams are in use. If multiple streams are created simultaneously, one or two streams could remain a dedicated command and control stream while the others are used for application data.

    \subsection{Mutual Authentication}
    Mutual authentication offers an additional layer of security for scenarios where identity of the target server is critical.

    \subsection{Bot Net Mitigation}
    If USP is implemented in the operating system, the agent can act as a gatekeeper for processes attempting to initiate outbound connections or listen for new connections. The operating systems can implement policies to control the USP agent and determine whether a process should be allowed to create or listen for connections. The USP agent would be capable of throttling the number of connections a process is allowed to make over a given time period, significantly reducing the effectiveness of rogue applications.

    \subsection{Redirects}
    A message could be formatted that provides information to the client for a third server so that the client will make a subsequent initialization request to a different host. This could be done before or after authentication. 
    
    Using the existing transport layer port scheme, this could be useful for having an agent listen to incoming connections and route them to different ports on the same server, enabling all applications to target a single port. When mature enough, USP could enhance the existing internet infrastructure by enabling servers to communicate with firewalls and dynamically configure port and IP rules, based on what was negotiated on the primary host port serving a USP agent operating only in a redirect capacity.

    \subsection{Presentation Layer}
    Although traditionally not conceptualized as a function of the presentation layer, encryption schemes could be negotiated by USP for each stream independently of transport or application layer protocols.

    The presentation layer may also provide some benefit to dynamically load processing capabilities for a particular application. If an application is specified in a URL that the client application does not have processing capabilities for, a presentation layer protocol could load a library, such as a DLL, to use in an extensible application protocol processing system. 

    \subsection{Establishing Multiple Streams Simultaneously}
    Establishing multiple streams for the same application session may be beneficial by giving more flexibility to how data is segmented and multiplexed between client and server. The "initialize" and "token" messages are constructed so that multiple streams may be declared and each stream may be authenticated separately. If USP is implemented in the operating system, and the exchange of port numbers is revised, it would be possible to establish multiple simultaneous sessions by simply having the client specify a port number and a quantity of sequential ports to be established. The server will establish the same number of streams, and then pass those streams to the USP agent to negotiate how many, if any, will be used, what they will be used for, and what to do if there are too many or not enough for the particular application.
    
    To complete each connection and ensure that streams are serving a matching purpose on the client and server, the server will need to send "connect" messages over each stream individually, specifying the application for that stream. The stream used to send the "initialize" message will be used for all USP session management messages, such as "initialize", "authenticate", "token", and "error". When the first connect is received on the primary command stream, the agents will need to iterate through the additional streams algorithmically, using "connect" and "error" message to complete or terminate each stream individually. At the end of this process, the client and server should have the same number of streams connected, with matching application purposes on each system, and use program logic to determine what, if anything, to do about streams that did not connect.

    \subsection{Electronic Voting from Personal Devices}
    Democratic voting systems seek to balance two conflicting and mutually exclusive principles: anonymity and identity. Establishing identity is important to ensure that only legitimate stakeholders are participating in the vote; so long as there is an in-group/out-group division, we want to ensure that only members of the in-group are voting. Anonymity is important because it protects the integrity of individual choice and prevents political retribution. In physical space, we can accomplish this by creating an identity based entry system to a contained anonymous process: an ID is checked at the door, then an anonymous ballot is cast. With the current TCP/IP model, this is impossible. Applications must check every submission for identity in order to ensure that only legitimate submissions are processed. Attempting to authenticate and then allow an anonymous ballot would be equivalent to allowing unrestricted anonymous ballots. By contrast, the USP model enables replication of the physical model in application function: an identity based session can be established and the application can process the voting data anonymously. For most use cases, when the USP stream is returned, it can contain a property with an identity token that can be used to correlate application activity with an identity for accountability purposes. For a voting system, the USP stream would simply not contain a property for identity information. In that way, the vote recording application will never have access to identity, rendering it impossible to record an identity with a vote, ensuring that a known legitimate user has cast an anonymous ballot.

    While this does require that the system is designed honestly to achieve these goals, USP enables a novel application that is not possible with the existing TCP/IP model.

  \section{Limitations}
    \subsection{Application Naming}
    If USP is implemented in existing applications using the current TCP port targeting scheme, there would be no naming conflicts because each port would serve a unique set of applications. When implementing in the operating system, because applications are identified exclusively by their application name, the name must be unique for that host. This will be a challenge for some applications. A common pattern for systems is to host a primary HTTP service on a standard port and a secondary, administrative HTTP service on a non-standard port. For USP, this would cause a collision of application names. This could potentially be circumvented by giving the administrative service a unique application name, such as HTTP\_ADMIN. However, that could, theoretically, lead to an absurd proliferation of custom application names. 

    For machines hosting multiple services, there is a potential for application name conflicts in general; running two HTTP servers hosting different applications will cause a conflict. As virtual infrastructure becomes the norm for application providers, the trend has been and continues to be toward increasingly isolating application functions to smaller work units. That is, today, it is more common to run a single application on a virtual machine rather than the approach of hosting numerous applications with different purposes on one physical machine. Even the individual functions running in modern cloud infrastructure make use of short lived, single purpose, one time use, ephemeral virtual machines as their underlying platform, thereby avoiding naming conflicts entirely. From this perspective, it seems increasingly less likely that any given machine will experience a naming conflict.

    An additional problem is that the client and server application need to both understand what it means to process an application by a given name. A web browser may be able to establish a stream with a server hosting HTTP\_ADMIN, but the only way for communication to work is if the browser understands that HTTP\_ADMIN data should be processed as HTTP data. There may be some approach to resolving this if it's possible to create another protocol that can provide the necessary information for the client application to adapt to the server application. For example, a web browser that access an HTTP\_ADMIN application which it was not designed to process could be directed to a repository of extensions that enables HTTP\_ADMIN processing, or simply informed that HTTP\_ADMIN data should be processed as HTTP data. This may be what the original presentation layer could have accomplished.

    \subsection{Adoption in Existing Infrastructure}
    In either application or operating system implementations, adoption of USP is incompatible with the existing TCP/IP model because adding a session layer is a change to the architecture of TCP/IP itself. When developing applications, one possible mitigation for this would be for the application server to simply ignore malformed messages, i.e. messages relevant to application layer protocols, and then to continue listening for USP messages. In this case, the client application could first attempt to connect using USP, and if no response is received, then attempt to connect to the application layer protocol itself. Application servers implementing USP would also know what USP and application layer messages to expect, so they would be able to determine whether to route application layer traffic directly to the application or process USP messages, and whether to strictly require USP or permit application layer directly, perhaps while transitioning to a new TCP/IP model.

    Adopting the TCP port exchange scheme and implementation of USP in the operating system are decidedly more complex and resistant to change, as an existing design element of the current TCP/IP model, and all libraries built on that model, would need to be revised. It may be possible to design the TCP layer to accommodate either connection request format, but this will require significant research.

    \bibliography{usp}
\end{document}